\documentstyle{mn}
\newif\ifAMStwofonts
\input psfig.tex
\ifoldfss
  \ifCUPmtlplainloaded \else
    \NewTextAlphabet{textbfit} {cmbxti10} {}
    \NewTextAlphabet{textbfss} {cmssbx10} {}
    \NewMathAlphabet{mathbfit} {cmbxti10} {} % for math mode
    \NewMathAlphabet{mathbfss} {cmssbx10} {} %  "   "    "
  \fi
  \ifAMStwofonts
    \ifCUPmtlplainloaded \else
      \NewSymbolFont{upmath} {eurm10}
      \NewSymbolFont{AMSa} {msam10}
      \NewMathSymbol{\upi}     {0}{upmath}{19}
      \NewMathSymbol{\umu}     {0}{upmath}{16}
      \NewMathSymbol{\upartial}{0}{upmath}{40}
      \NewMathSymbol{\leqslant}{3}{AMSa}{36}
      \NewMathSymbol{\geqslant}{3}{AMSa}{3E}

    \fi
  \fi
\fi % End of OFSS

\ifnfssone
  \newmathalphabet{\mathit}
  \addtoversion{normal}{\mathit}{cmr}{m}{it}
  \addtoversion{bold}{\mathit}{cmr}{bx}{it}
  \newmathalphabet{\mathbfit} % math mode version of \textbfit{..}
  \addtoversion{normal}{\mathbfit}{cmr}{bx}{it}
  \addtoversion{bold}{\mathbfit}{cmr}{bx}{it}
  \newmathalphabet{\mathbfss} % math mode version of \textbfss{..}
  \addtoversion{normal}{\mathbfss}{cmss}{bx}{n}
  \addtoversion{bold}{\mathbfss}{cmss}{bx}{n}
  \ifAMStwofonts
    \ifCUPmtlplainloaded \else
      \UseAMStwoboldmath
      \makeatletter
      \new@mathgroup\upmath@group
      \define@mathgroup\mv@normal\upmath@group{eur}{m}{n}
      \define@mathgroup\mv@bold\upmath@group{eur}{b}{n}
      \edef\UPM{\hexnumber\upmath@group}
      \new@mathgroup\amsa@group
      \define@mathgroup\mv@normal\amsa@group{msa}{m}{n}
      \define@mathgroup\mv@bold\amsa@group{msa}{m}{n}
      \edef\AMSa{\hexnumber\amsa@group}
      \makeatother
      \mathchardef\upi="0\UPM19
      \mathchardef\umu="0\UPM16
      \mathchardef\upartial="0\UPM40
      \mathchardef\leqslant="3\AMSa36
      \mathchardef\geqslant="3\AMSa3E
    \fi
  \fi
\fi % End of NFSS release 1

\ifnfsstwo
  \DeclareMathAlphabet{\mathbfit}{OT1}{cmr}{bx}{it}
  \SetMathAlphabet\mathbfit{bold}{OT1}{cmr}{bx}{it}
  \DeclareMathAlphabet{\mathbfss}{OT1}{cmss}{bx}{n}
  \SetMathAlphabet\mathbfss{bold}{OT1}{cmss}{bx}{n}
  \ifAMStwofonts
    \ifCUPmtlplainloaded \else
      \DeclareSymbolFont{UPM}{U}{eur}{m}{n}
      \SetSymbolFont{UPM}{bold}{U}{eur}{b}{n}
      \DeclareSymbolFont{AMSa}{U}{msa}{m}{n}
      \DeclareMathSymbol{\upi}{0}{UPM}{"19}
      \DeclareMathSymbol{\umu}{0}{UPM}{"16}
      \DeclareMathSymbol{\upartial}{0}{UPM}{"40}
      \DeclareMathSymbol{\leqslant}{3}{AMSa}{"36}
      \DeclareMathSymbol{\geqslant}{3}{AMSa}{"3E}
    \fi
  \fi
\fi % End of NFSS release 2

\ifCUPmtlplainloaded \else
  \ifAMStwofonts \else % If no AMS fonts
    \def\upi{\pi}
    \def\umu{\mu}
    \def\upartial{\partial}
  \fi
\fi

\title[Mn {\sc i} HFS in the near-infrared]{Mn {\sc i} hyperfine structure in the near-infrared}
\author[J. Mel\'endez]
       {Jorge Mel\'endez \thanks{Affiliated to Seminario Permanente de Astronomia y 
        Ciencias Espaciales, Universidad Nacional Mayor de San Marcos, Per\'u. E-mail: jorge@iagusp.usp.br} \\
Universidade de S\~ao Paulo, Cx.P. 3386, S\~ao Paulo - SP, 01060-970, Brazil}
\date{Accepted ...
      Received ...}
\pagerange{\pageref{firstpage}--\pageref{lastpage}}
\pubyear{1999}

\begin{document}

\maketitle

\label{firstpage}

\begin{abstract}
Hyperfine interaction constants and hyperfine components of \hbox{Mn\,{\sc i}}
lines in the near-infrared {\it J} and {\it H} bands were obtained by fitting the solar 
spectrum. I identified the 17744-{\AA} solar absorption line as resulting from
\hbox{Mn\,{\sc i}}, and I discarded the identifications of the 13281.65-{\AA} 
and 16929.85-{\AA} solar features as a result of  \hbox{Fe\,{\sc i}} and 
\hbox{Mn\,{\sc i}}, respectively.
\end{abstract}


\section{Introduction}

{Mn {\sc i}} shows a large hyperfine structure (HFS) owing to the interaction 
of the electronic ($\bmath J$) and nuclear ($\bmath I$) angular momenta. The
complex profiles have been studied mainly in the ultraviolet and optical 
regions but not extensively in the infrared region. In the {\it J} (1.00 - 1.34
$\mu$m) and {\it H} (1.49 - 1.80 $\mu$m) bands of the solar spectrum there are 
several \hbox{Mn\,{\sc i}} lines (Livingston \& Wallace 1991) which show 
clearly wide hyperfine profiles.

Abt (1952) showed that the profiles and strengths of \hbox{Mn\,{\sc i}} 
stellar absorption lines depend on the HFS. Booth \& Blackwell (1983) found 
that even for weak lines errors of 0.1 dex are possible when HFS is neglected. 
Hence, it is very important to consider HFS for accurate abundance analysis. 
The most extensive laboratory work to measure HFS of \hbox{Mn\,{\sc i}} 
lines was carried out by Booth, Shallis \& Wells (1983), but unfortunately 
their measurements only cover ultraviolet and optical lines. 

I intend to carry out near-infrared (1-2 $\mu$m) spectroscopic 
observations of metal-rich stars in the Galactic bulge.  The infrared 
region offers the best option to study these very reddened stars because 
of the high extinction in the optical region. Owing to the lack of HFS data 
for the {Mn {\sc i}} infrared lines, I decided to obtain the HFS components 
in order to carry out accurate spectroscopic work 
in the {\it J} and {\it H} bands.

In Section 2 the hyperfine components are derived, in Section 3 the 
spectrum synthesis calculations are described and finally the summary 
of the results is presented in Section 4.

\section{Hyperfine components}

The energy hyperfine splittings ($\Delta E_F$) were obtained from the 
magnetic dipole ($A$) and electric quadrupole ($B$) interaction 
constants (e.g. Emery 1996):

\begin{equation}
\Delta E_F= \frac{1}{2} AC+B
\frac{\frac{3}{4} C(C+1) -
I(I+1)J(J+1)}{2I(2I-1)J(2J-1)},
\end{equation}

where $C=F(F+1)-J(J+1)-I(I+1)$ and
$F=J+I,J+I-1,...,|J-I|$. For Mn the nuclear spin is $I$ = 5/2.

The interaction constants were taken from  White \& Ritschl (1930), 
Handrich, Streudel \& Walther (1969), Luc \& Gerstenkorn (1972), 
Beynon (1977), Dembczynski et al. (1979), Kronfeldt et al. (1985), 
Brodzinski et al. (1987), the unpublished results of Guthr\"oehrlein 
(as given in Brodzinski et al. 1987), and the theoretical values calculated 
by Per J\"onsson (private communication). J\"onsson's values are shown in 
column 3 of Table 11, they were kindly calculated by him using the 
GRASP relativistic code (J\"onsson, Parpia \& Froese Fischer 1996).

From the splittings levels ($\Delta E_F$) I computed the wavenumbers of 
the components using the selection rules $\Delta F$ = -1, 0, +1. The air 
wavelengths were obtained from the wavenumbers using the dispersion formula 
of Edl\'en (1966). The relative intensities of the components, $S_{FF'}$, 
were calculated using the formula (e.g. Emery 1996) 

\begin{equation}
S_{FF'} \; = \; (2F+1)(2F'+1) \left\{ \begin{array}{lll} F & F' & 1 \\
                                J' & J & I  \end{array}
                               \right\}^2 ,
\end{equation}

the quantity in the braces \{ \} is a 6$j$ symbol. The 6$j$ symbol was 
calculated using a subroutine taken from Robert Cowan's atomic structure code.

Initially, the relative wavelength positions of the components were 
calculated with the interaction constants from the given references,
and then the interaction constants were adjusted to achieve the best
match with the solar spectrum, as described in detail in Section 3.
The final values of the relative positions and intensities of the components, 
with respect to the main (strongest) hyperfine component, are given in 
Tables 2-10  and displayed with vertical lines in Figs 1-3.

\section{Spectrum synthesis}

The code for spectrum synthesis was described by Barbuy (1981, 1982). The 
synthetic spectrum was calculated with the local thermodynamic equilibrium 
(LTE) approximation, in successive steps of 0.02 {\AA} in wavelength
and convolved with an instrumental profile. A solar model, interpolated in 
the unpublished grids of model atmospheres by B. Gustafsson, was adopted. 
The convective fluxes of Gustafsson's models were evaluated using the variant 
of the mixing-length theory given by Henyey, Vardya \& Bodenheimer (1965). 
Details of the model atmospheres are given by Gustafsson et al. (1975). 
The adopted solar abundances were those reported by Grevesse, 
Noels \& Sauval (1996), $A_{Mn}$ = 5.39 $\pm$ 0.03 dex.

It is important to consider collisional broadening by hydrogen atoms in 
the \hbox{Mn\,{\sc i}} lines because it could affect the determination 
of hyperfine constants from the solar spectrum. The van der Waals line 
broadening is given by $\gamma_6/N_{\rm H} = 17 v^{3/5} C_6^{2/5}$
(Allen 1955) where $v$ is velocity, $N_{\rm H}$ is the number density of 
hydrogen and $C_6$ is the interaction constant for collision broadening. 
$C_6$ was computed by using the tables of accurate line broadening
cross-sections $\sigma$ of Anstee \& O'Mara (1995), Barklem \& O'Mara (1997) 
and Barklem, O'Mara \& Ross  (1998), for transitions between states 
s-p/p-s, p-d/d-p, and d-f/f-d, respectively. These tables give $\sigma$ in 
terms of the effective principal quantum numbers ($n^*$) of the upper and 
lower states of the transition. The obtained $C_6$ values for 
the \hbox{Mn\,{\sc i}} lines are given in Table 1 among other atomic data. 

Atomic (mainly Fe) and CN lines, blended with the \hbox{Mn\,{\sc i}} lines, 
were also included in the calculations. Details about the molecular and 
atomic infrared line lists are given in Mel\'endez \& Barbuy (1999).

The solar identifications of the \hbox{Mn\,{\sc i}} lines were taken 
from Livingston \& Wallace (1991), unless it was established to the contrary. 
The solar spectrum considered for the comparison with the synthetic 
spectrum was that given by Livingston \& Wallace (1991).

\subsection{The 12899-, 13293-, 13319- and 17744-{\AA} lines}
These lines were well reproduced with the laboratory interaction constants 
given in the literature, and I only needed adjust the oscillator 
strength empirically until the synthetic spectrum matched the solar spectrum.
However, it was necessary to correct wavelength positions using the 
solar spectrum, not only for these lines, but also for all the lines in 
this paper. Taklif (1990) found differences between measured and 
predicted wavenumbers of fine structure components. He suggested further 
work on checking the accuracy of some energy levels.

In Fig. 1 the fittings are shown and in Table 1 the obtained 
solar wavelengths and log{\it (gf)} values for the main hyperfine 
components are given. The wavelength positions and oscillator strengths of the 
others HFS components could be obtained using the relative positions and 
intensities given in Tables 2-10. From the fittings, I estimated that 
the errors for the {\it gf} values, given in Table 1, are less than 0.05 dex
(without including the error in the assumed abundance).

The 12899- and 13319-{\AA} lines have central depths 
($d \; = \; 1 \; - \; I_{line}/I_{cont}$) of about 0.5 and are affected 
by NLTE effects, for this reason the observed profile is slightly deeper 
than the synthetic one.

The 17744-{\AA} line was identified by myself using the laboratory line 
list of Taklif (1990). This line shows the widest separation between 
hyperfine components (2 {\AA}). Some part of the solar spectrum that is 
affected by telluric absorption was not recovered, but the fitting 
was satisfactory.

\subsection{The 12976- and 13281-{\AA} lines}

In the case of the 12976- and 13281-\AA\ lines (Fig. 2), I needed to 
make a correction to the experimental $A$ hyperfine interaction constants
 in order to reproduce the solar line profiles. The upper levels 
laboratory $A$ values are highly accurate and therefore were maintained fixed 
while the lower levels $A$ values  were adjusted until the synthetic 
spectrum matched the solar spectrum. 

The most work was required to fit the the 13281-{\AA} line (Fig. 2). In 
the atlas of Livingston \& Wallace (1991) this \hbox{Mn\,{\sc i}} line 
is shown to be blended with a strong \hbox{Fe\,{\sc i}} line 
at 13281.65 {\AA}. This \hbox{Fe\,{\sc i}} line was not included in the 
recent \hbox{Fe\,{\sc i}} multiplet table of Nave et al. (1994) and 
therefore I did not have available the excitation potential for this line. 
In a first attempt, I assigned a value of 5.5 eV and the line was 
well reproduced with log{\it (gf)} = -0.99. However, I could not fit 
simultaneously the empirical \hbox{Fe\,{\sc i}} and the \hbox{Mn\,{\sc i}} 
lines. So, I thought that the \hbox{Fe\,{\sc i}} line would not be the 
main contributor to the 13281.65-{\AA} solar feature, and I discarded it 
and tried to fit all the profile as being due only to 
the \hbox{Mn\,{\sc i}} HFS. Then after some adjustment to the $A$ value of 
the lower level, I obtained the profile shown in Fig. 2 (lower panel).

Livingston \& Wallace (1991) used the solar identifications given by 
Bi\'emont et al. (1985) to assign the 13281.65-{\AA} solar feature 
to \hbox{Fe\,{\sc i}}, but the laboratory intensity of that line is 
weak (Bi\'emont et al. 1985) and hardly it would be the main contributor 
to that solar absorption line. I suggest that the 13281.65-{\AA} solar 
feature is mainly a result of \hbox{Mn\,{\sc i}}, specifically to 
the {\it F} = 5 to {\it F'} = 4 hyperfine component.

In Table 11 the solar $A$ values obtained in this work are compared with 
the experimental values of White \& Ritschl (1930) and Beynon (1977). 
Despite the fact I used incompletely resolved line patterns, the solar 
values are reliable because they were obtained using accurate level 
splittings for the upper levels given by Kronfeldt et al. (1985). 
The solar $A$ values agree better with the earlier determination of 
White \& Ritschl (1930) than with the suggested values given by 
Beynon (1977). The estimated uncertainties of the solar $A$ values, as 
deduced from the solar fits, are given between parenthesis in Table 11. 
$B$ interaction constants could not be obtained from the solar fits, 
but their contribution to the splitting is usually negligible.

\subsection{The 15159-, 15217- and 15262-{\AA} lines}

Even though $A$ interaction constants of the lower levels of these lines 
are well known, there are no laboratory measurements of the $A$ values for 
the upper levels. In a first attempt to reproduce the solar profiles, 
the theoretical $A$ values of J\"onsson (Table 11) were employed. 
The theoretical factors were slightly varied until the best match 
between synthetic and solar profiles was obtained. In Table 11 the solar
$A$ values are compared with the theoretical ones.

In Fig. 3 the fittings are displayed. The blends at 15159.7 and 
15216.9 {\AA} are due to \hbox{Fe\,{\sc i}}.

\subsection{Other \hbox{Mn\,{\sc i}} lines probably present in the Sun}

Other \hbox{Mn\,{\sc i}} infrared lines which are probably present in 
the Sun are described here. 

The 11613-{\AA} line (e$^6$S$_{5/2}$ - v$^6$P$_{7/2}$) was identified in the 
Sun by Swensson et al. (1973). This very weak line is affected by telluric 
absorption and it was not possible to perfom a reliable fitting.  I found 
log{\it (gf)} $\la$ -0.40 dex for the main hyperfine component of this line.

In the list of the solar spectrum lines by Ramsauer, Solanki \& Bi\'emont 
(1995) the 16929.85-{\AA} line (e$^4$P$_{3/2}$ - 9p $^8$P$_{5/2}$) was 
identified as owing to \hbox{Mn\,{\sc i}}. Despite the fact that the 
theoretical oscillator strength given by Kurucz (1995) is very small 
[log({\it gf)} = -3.2] and the excitation potential relatively high (6.4 eV), 
I tried to fit this line. It was necessary to increase by about four 
orders of magnitude the theoretical oscillator strength in order to fit 
the solar line depth. I also noted that it was impossible to fit the 
overall profile of that line using the theoretical interaction constants 
of J\"onsson ($A$ = -0.001 cm$^{-1}$ and $B$ = 0.003 cm$^{-1}$ for 
the e$^4$P$_{3/2}$ state, and $A$ = 0.025 cm$^{-1}$ for 9p $^8$P$_{5/2}$). 
Even with reasonable variations in the $A$ values, it was not possible to fit 
the profile. So, I suggest that the identification of that feature, given 
by Ramsauer et al. (1995), could be wrong.

Using the laboratory line list of Taklif (1990) I identified the 
15967 (e$^8$D$_{11/2}$ - z$^8$F$_{13/2}$), 16707 (e$^6$S$_{5/2}$ - w$^6$P$_{7/2}$), 
17325 (e$^6$D$_{9/2}$ - w$^6$F$_{11/2}$), 17339 (e$^6$D$_{7/2}$ - w$^6$F$_{9/2}$) and 
17349 (e$^6$D$_{5/2}$ - w$^6$F$_{7/2}$) {\AA} lines in the solar spectrum, but 
they are severely blended, most of them with unknown contributors, therefore 
it was not possible to perfom reliable fittings for these lines.

\section{SUMMARY}

Applying spectrum synthesis calculations I reproduced successfully the
hyperfine pattern of the \hbox{Mn\,{\sc i}} near infrared lines present 
in the {\it J} and {\it H} bands of the solar spectrum.  Using the solar spectrum,
new values for five hyperfine interaction constants were obtained.

According to the results given in this work the principal contributor 
to the 13281.65-{\AA} solar feature is the main hyperfine 
component {\it F} = 5 to {\it F'} = 4 and not \hbox{Fe\,{\sc i}}. 
On the other hand, I discard the identification of the 16929.85-{\AA} solar 
line as resulting from \hbox{Mn\,{\sc i}}.

The relative positions and intensities of the hyperfine components were 
given, and also the oscillator strength and wavelength of the principal 
component. So, I have obtained all the information that I need 
about \hbox{Mn\,{\sc i}} before starting to analyse accurately the spectrum 
of other cool stars in the near infrared region. However, 
laboratory measurements are necessary to improve the accuracy of the 
energy levels and to determine more accurate HFS splittings.

\bigskip
\section*{Acknowledgements}

The calculations were carried out on a DEC Alpha 300/700 workstation 
provided by FAPESP. The author thanks support by FAPESP fellowship 97/0109-8. 
He is also grateful to Per J\"onsson for the calculation of the 
theoretical hyperfine interaction constants. NSO/Kitt Peak FTS data used 
here were produced by NSF/NOAO.

\newpage
\begin{table}
\centering
 \begin{minipage}{130mm}
\caption{\hbox{Mn\,{\sc i}} atomic lines in the {\it J} and {\it H} bands \label{tabla1}}
\begin{tabular}{ccccc}
\hline \hline
{Wavelength}\footnote{Wavelength for the main hyperfine component as obtained 
from \\ the solar spectrum} & {Transition} &
{$\chi$} \footnote{Excitation energy for the lower level (Kurucz 1995)}  &
{log{\it (gf)}} \footnote{Value determined in this work for the main 
hyperfine component} & {$C_6$ ($10^{-31}$} \\
{({\AA})} &         {}     & {(eV)} & {} & {cm$^{6}$ s$^{-1}$)} \\
\hline
12899.87& a$^6$D$_{9/2}$ - z$^6$P$_{7/2}$ & 2.114 & -1.76 & 0.21 \\
12976.05& a$^4$D$_{7/2}$ - z$^4$P$_{5/2}$ & 2.889 & -1.57 & 0.20 \\
13281.63& a$^4$D$_{5/2}$ - z$^4$P$_{3/2}$ & 2.920 & -1.80 & 0.20 \\
13293.82& a$^6$D$_{7/2}$ - z$^6$P$_{7/2}$ & 2.143 & -2.27 & 0.21 \\
13319.06& a$^6$D$_{7/2}$ - z$^6$P$_{5/2}$ & 2.143 & -2.00 & 0.21 \\
15159.21& e$^8$S$_{7/2}$ - y$^8$P$_{9/2}$ & 4.889 & -0.12 & 5.49 \\
15217.76& e$^8$S$_{7/2}$ - y$^8$P$_{7/2}$ & 4.889 & -0.19 & 5.46 \\
15262.50& e$^8$S$_{7/2}$ - y$^8$P$_{5/2}$ & 4.889 & -0.28 & 5.44 \\
17743.58& y$^6$P$_{7/2}$ - e$^6$S$_{5/2}$ & 4.435 & -1.37 & 3.94 \\
\hline \hline
\end{tabular}
\end{minipage}
\end{table}

\begin{table}
\begin{minipage}{140mm}
\caption{Hyperfine components (12899 {\AA})}
\begin{tabular}{r r r}
\hline \hline
{$\Delta \lambda$}  & {relative} & {\it F - F'} \\
{({\AA})} & {intensities} & {} \\
\hline
-0.374 &   0.3 & 5 - 6 \\
-0.371 &   0.7 & 4 - 5 \\
-0.364 &   0.8 & 3 - 4 \\
-0.354 &   0.6 & 2 - 3 \\
-0.283 &   7.7 & 2 - 2 \\
-0.270 &  11.9 & 3 - 3 \\
-0.252 &  13.5 & 4 - 4 \\
-0.237 &  25.0 & 2 - 1 \\
-0.230 &  12.4 & 5 - 5 \\
-0.203 &   8.0 & 6 - 6 \\
-0.200 &  34.0 & 3 - 2 \\
-0.158 &  45.8 & 4 - 3 \\
-0.111 &  60.7 & 5 - 4 \\
-0.058 &  78.6 & 6 - 5 \\
 0.000 & 100.0 & 7 - 6 \\
\hline \hline
\end{tabular}
\end{minipage}
\end{table}

\begin{table}
 \begin{minipage}{140mm}
\caption{Hyperfine components (12975 {\AA})}
\begin{tabular}{r r r}
\hline \hline
{$\Delta \lambda$}  & {relative} & {\it F - F'} \\
{({\AA})} & {intensities} & {} \\
\hline
-0.332 &  10.3 & 1 - 0 \\
-0.314 &  19.8 & 2 - 1 \\
-0.296 &  11.0 & 1 - 1 \\
-0.271 &  33.0 & 3 - 2 \\
-0.244 &  16.5 & 2 - 2 \\
-0.226 &   1.8 & 1 - 2 \\
-0.203 &  50.4 & 4 - 3 \\
-0.167 &  19.2 & 3 - 3 \\
-0.140 &   2.2 & 2 - 3 \\
-0.113 &  72.5 & 5 - 4 \\
-0.066 &  18.1 & 4 - 4 \\
-0.030 &   1.6 & 3 - 4 \\
 0.000 & 100.0 & 6 - 5 \\
 0.056 &  12.1 & 5 - 5 \\
 0.102 &   0.7 & 4 - 5 \\
\hline \hline
\end{tabular}
\end{minipage}
\end{table}

\begin{table}
 \begin{minipage}{140mm}
\caption{Hyperfine components (13281 {\AA})}
\begin{tabular}{r r r}
\hline \hline
{$\Delta \lambda$}  & {relative} & {\it F - F'} \\
{({\AA})} & {intensities} & {} \\
\hline
-0.328  & 12.7 & 2 - 1 \\
-0.311  & 19.1 & 1 - 1 \\
-0.302  &  9.1 & 0 - 1 \\
-0.260  & 32.7 & 3 - 2 \\
-0.235  & 27.3 & 2 - 2 \\
-0.218  &  8.2 & 1 - 2 \\
-0.151  & 61.4 & 4 - 3 \\
-0.118  & 28.6 & 3 - 3 \\
-0.093  &  5.5 & 2 - 3 \\
 0.000  &100.0 & 5 - 4 \\
 0.042  & 20.5 & 4 - 4 \\
 0.075  &  2.3 & 3 - 4 \\
\hline \hline
\end{tabular}
\end{minipage}
\end{table}

\begin{table}
\begin{minipage}{140mm}
\caption{Hyperfine components (13293 {\AA})}
\begin{tabular}{r r r}
\hline \hline
{$\Delta \lambda$}  & {relative} & {\it F - F'} \\
{({\AA})} & {intensities} & {}  \\
\hline
-0.164 &  10.2 & 5 - 6 \\
-0.147 &  15.5 & 4 - 5 \\
-0.130 &  16.7 & 3 - 4 \\
-0.111 &  14.4 & 2 - 3 \\
-0.091 &   9.1 & 1 - 2 \\
-0.042 &  16.3 & 1 - 1 \\
-0.036 &  19.0 & 2 - 2 \\
-0.029 &  28.3 & 3 - 3 \\
-0.020 &  44.1 & 4 - 4 \\
-0.011 &  67.5 & 5 - 5 \\
 0.000 & 100.0 & 6 - 6 \\
 0.013 &   9.1 & 2 - 1 \\
 0.045 &  14.4 & 3 - 2 \\
 0.080 &  16.7 & 4 - 3 \\
 0.116 &  15.5 & 5 - 4 \\
 0.154 &  10.2 & 6 - 5 \\
\hline \hline
\end{tabular}
\end{minipage}
\end{table}

\begin{table}
\begin{minipage}{140mm}
\caption{Hyperfine components (13319 {\AA})}
\begin{tabular}{r r r}
\hline \hline
{$\Delta \lambda$}  & {relative} & {\it F - F'} \\
{({\AA})} & {intensities} & {}  \\
\hline
-0.295 &   0.7 & 4 - 5 \\
-0.266 &   1.6 & 3 - 4 \\
-0.235 &   2.2 & 2 - 3 \\
-0.202 &   1.8 & 1 - 2 \\
-0.161 &  12.1 & 5 - 5 \\
-0.159 &  18.1 & 4 - 4 \\
-0.155 &  19.2 & 3 - 3 \\
-0.150 &  16.5 & 2 - 2 \\
-0.145 &  11.0 & 1 - 1 \\
-0.117 &  10.3 & 1 - 0 \\
-0.093 &  19.8 & 2 - 1 \\
-0.071 &  33.0 & 3 - 2 \\
-0.048 &  50.4 & 4 - 3 \\
-0.025 &  72.5 & 5 - 4 \\
 0.000 & 100.0 & 6 - 5 \\
\hline \hline
\end{tabular}
\end{minipage}
\end{table}

\begin{table}
\begin{minipage}{140mm}
\caption{Hyperfine components (15159 {\AA})}
\begin{tabular}{r r r}
\hline \hline
{$\Delta \lambda$}  & {relative} & {\it F - F'} \\
{({\AA})} & {intensities} & {} \\
\hline
-0.183 &  34.0 & 2 - 3 \\
-0.183 &  25.0 & 1 - 2 \\
-0.166 &  45.8 & 3 - 4 \\
-0.129 &  60.7 & 4 - 5 \\
-0.074 &  78.6 & 5 - 6 \\
-0.070 &   7.7 & 2 - 2 \\
-0.014 &  11.9 & 3 - 3 \\
 0.000 & 100.0 & 6 - 7 \\
 0.061 &  13.5 & 4 - 4 \\
 0.100 &   0.6 & 3 - 2 \\
 0.154 &  12.4 & 5 - 5 \\
 0.212 &   0.8 & 4 - 3 \\
 0.265 &   8.0 & 6 - 6 \\
 0.343 &   0.7 & 5 - 4 \\
 0.493 &   0.3 & 6 - 5 \\
\hline \hline
\end{tabular}
\end{minipage}
\end{table}

\begin{table}
\begin{minipage}{140mm}
\caption{Hyperfine components (15217 {\AA})}
\begin{tabular}{r r r}
\hline \hline
{$\Delta \lambda$}  & {relative} & {\it F - F'} \\
{({\AA})} & {intensities} & {} \\
\hline
-0.349 &  15.5 & 4 - 5 \\
-0.345 &  16.7 & 3 - 4 \\
-0.342 &  10.2 & 5 - 6 \\
-0.331 &  14.4 & 2 - 3 \\
-0.306 &   9.1 & 1 - 2 \\
-0.213 &  16.3 & 1 - 1 \\
-0.192 &  19.0 & 2 - 2 \\
-0.160 &  28.3 & 3 - 3 \\
-0.117 &  44.1 & 4 - 4 \\
-0.099 &   9.1 & 2 - 1 \\
-0.064 &  67.5 & 5 - 5 \\
-0.021 &  14.4 & 3 - 2 \\
 0.000 & 100.0 & 6 - 6 \\
 0.068 &  16.7 & 4 - 3 \\
 0.168 &  15.5 & 5 - 4 \\
 0.278 &  10.2 & 6 - 5 \\
\hline \hline
\end{tabular}
\end{minipage}
\end{table}

\begin{table}
\begin{minipage}{140mm}
\caption{Hyperfine components (15262 {\AA})}
\begin{tabular}{r r r}
\hline \hline
{$\Delta \lambda$}  & {relative} & {\it F - F'} \\
{({\AA})} & {intensities} & {} \\
\hline
-0.630 &   0.7 & 4 - 5 \\
-0.539 &   1.6 & 3 - 4 \\
-0.455 &   2.2 & 2 - 3 \\
-0.378 &   1.8 & 1 - 2 \\
-0.344 &  12.1 & 5 - 5 \\
-0.310 &  18.1 & 4 - 4 \\
-0.283 &  19.2 & 3 - 3 \\
-0.263 &  16.5 & 2 - 2 \\
-0.249 &  11.0 & 1 - 1 \\
-0.185 &  10.3 & 1 - 0 \\
-0.135 &  19.8 & 2 - 1 \\
-0.091 &  33.0 & 3 - 2 \\
-0.054 &  50.4 & 4 - 3 \\
-0.024 &  72.5 & 5 - 4 \\
 0.000 & 100.0 & 6 - 5 \\
\hline \hline
\end{tabular}
\end{minipage}
\end{table}

\begin{table}
\begin{minipage}{140mm}
\caption{Hyperfine components (17744 {\AA})}
\begin{tabular}{r r r}
\hline \hline
{$\Delta \lambda$}  & {relative} & {\it F - F'} \\
{({\AA})} & {intensities} & {} \\
\hline
0.000 & 100.0 & 6 - 5 \\
0.246 &  12.1 & 5 - 5 \\
0.450 &   0.7 & 4 - 5 \\
0.670 &  72.5 & 5 - 4 \\
0.875 &  18.1 & 4 - 4 \\
1.039 &   1.6 & 3 - 4 \\
1.215 &  50.4 & 4 - 3 \\
1.379 &  19.2 & 3 - 3 \\
1.501 &   2.2 & 2 - 3 \\
1.633 &  33.0 & 3 - 2 \\
1.756 &  16.5 & 2 - 2 \\
1.838 &   1.8 & 1 - 2 \\
1.926 &  19.8 & 2 - 1 \\
2.008 &  11.0 & 1 - 1 \\
2.093 &  10.3 & 1 - 0 \\
\hline \hline
\end{tabular}
\end{minipage}
\end{table}

\begin{table}
\begin{minipage}{14cm}
\caption{Comparison of magnetic interaction constants ($A$)}
\begin{tabular}{@{}clrcc@{}}
\hline \hline
{Level} & {TW$_{\sun}$}\footnote{This work, from the solar spectrum} & 
          {J98}\footnote{P. J\"onsson, private communication} & 
          {WR30}\footnote{White \& Ritschl  (1930)} & 
          {B77}\footnote{Beynon (1977)} \\
{} & {(cm$^{-1}$)} & {(cm$^{-1}$)} & {(cm$^{-1}$)} & {(cm$^{-1}$)} \\
\hline
a$^4$D$_{7/2}$  & -0.0057 (0.0005)  &         & -0.0055 & -0.004 \\
a$^4$D$_{5/2}$  & -0.0046 (0.0005)  &         & -0.0042 & -0.003 \\
y$^8$P$_{9/2}$  & ~0.0165 (0.0015)  &  0.015  &         &        \\
y$^8$P$_{7/2}$  & ~0.0200 (0.0015)  &  0.018  &         &        \\
y$^8$P$_{5/2}$  & ~0.0275 (0.0015)  &  0.024  &         &        \\
\hline \hline
\end{tabular}
\end{minipage}
\end{table}

\newpage
\begin{figure}
{ {\psfig{figure=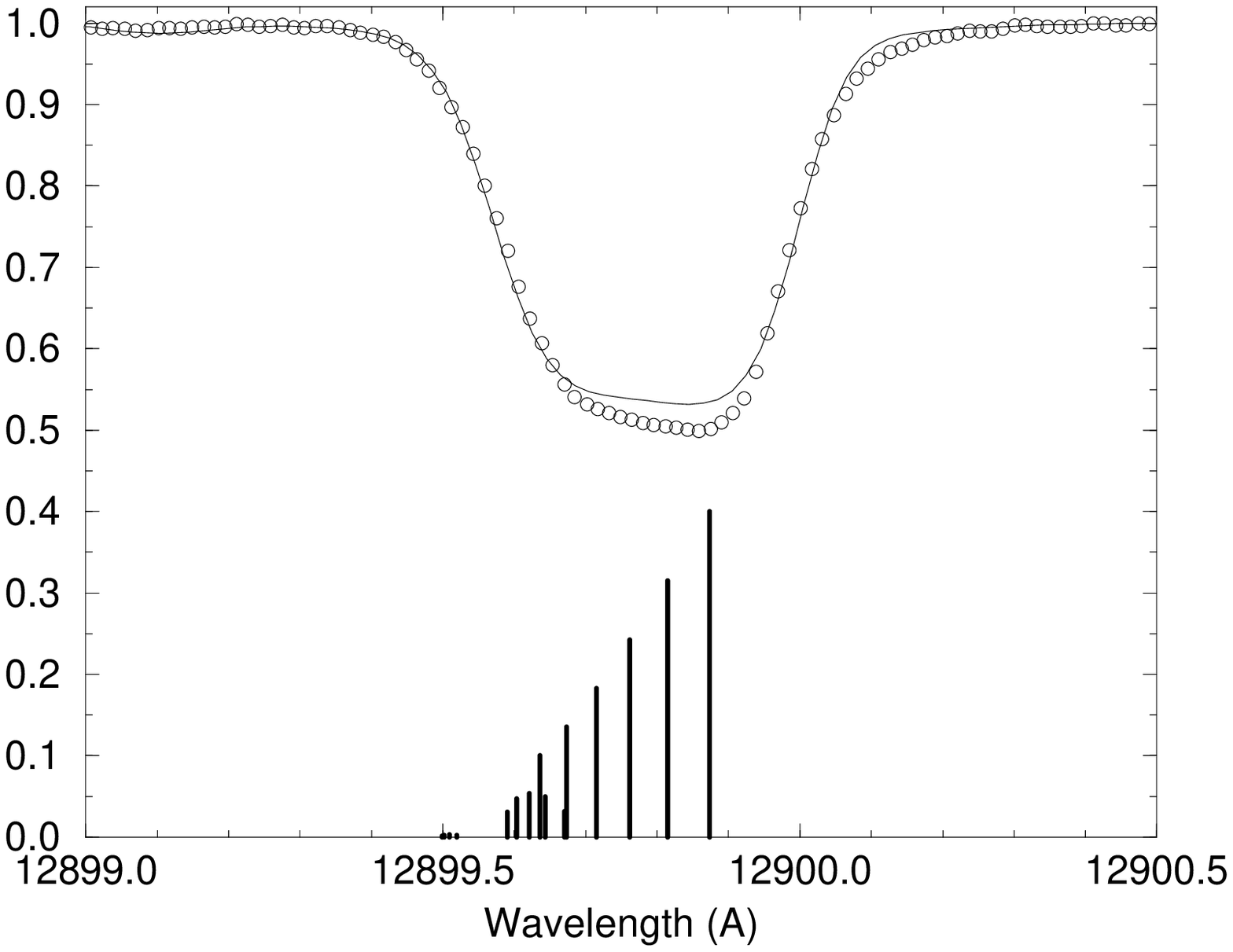,width=7cm}} {\psfig{figure=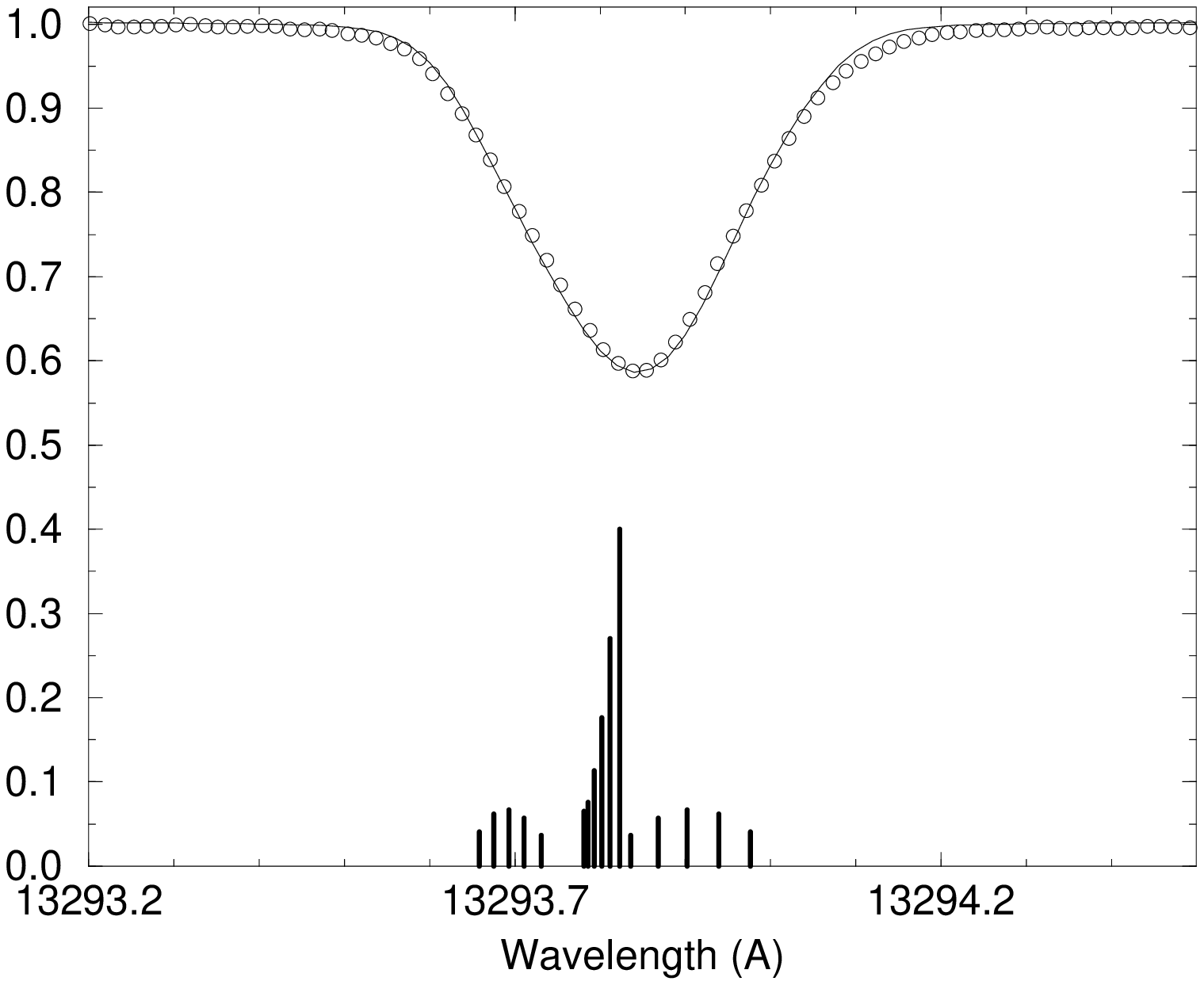,width=7cm}} 
{\psfig{figure=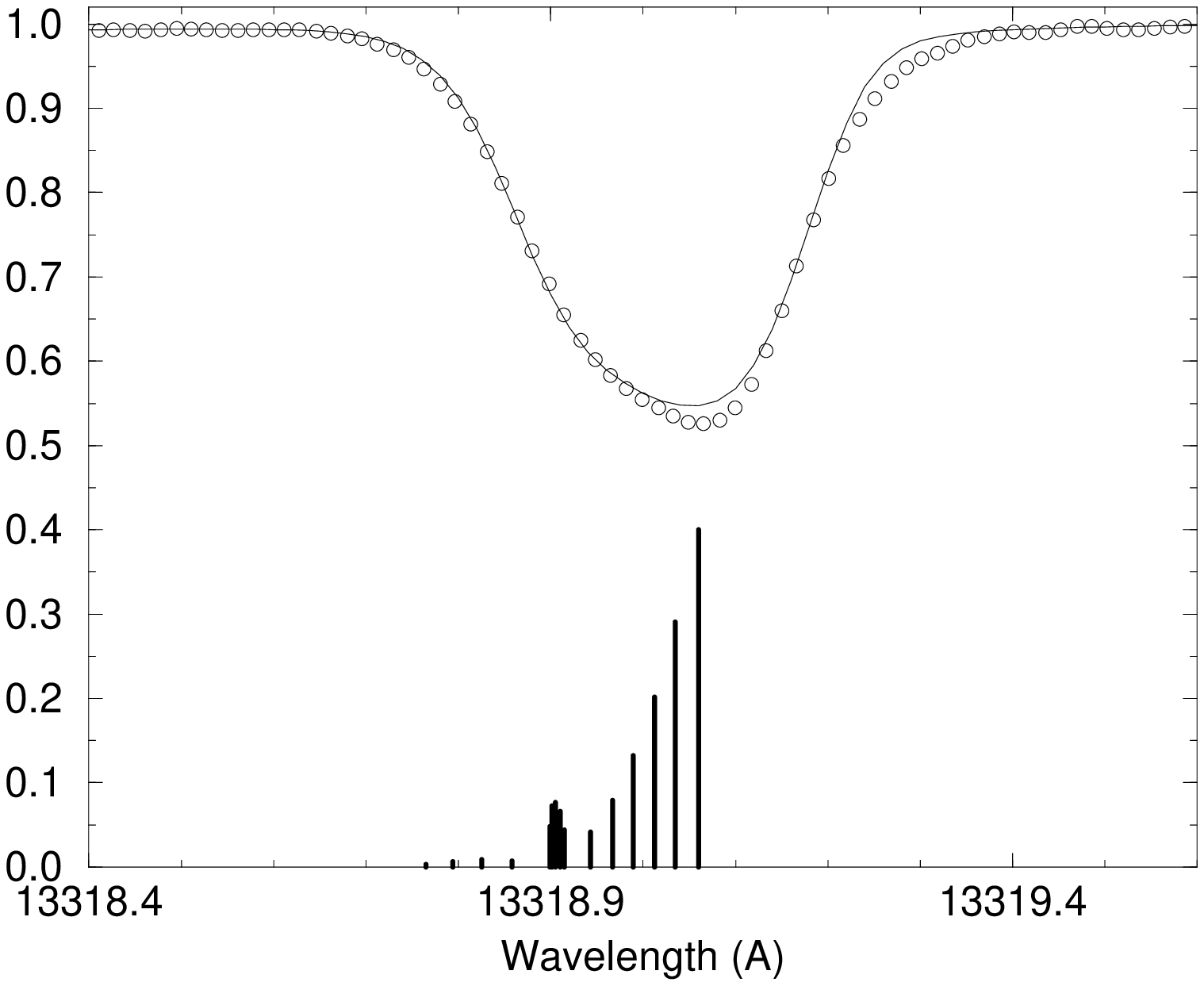,width=7cm}} {\psfig{figure=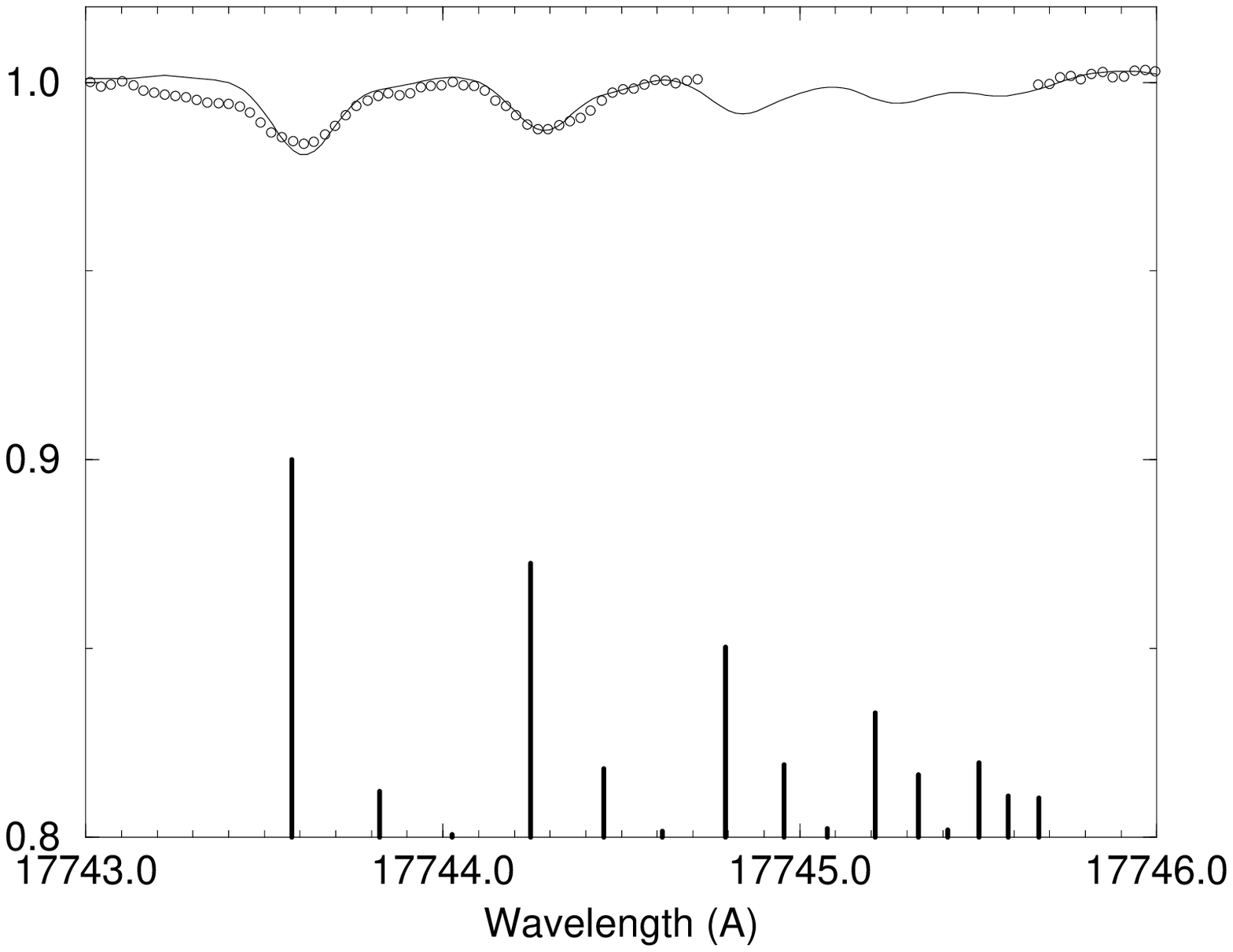,width=7cm}} }
\caption{Fittings for the 12899-, 13293-, 13319- and 17744-{\AA} lines. 
The HFS components were computed without any correction to laboratory interaction 
constants (Section 3.1). Solar spectrum (circles), synthetic spectrum 
(solid line) and HFS components (vertical lines).} 
\end{figure}

\begin{figure}
{ {\psfig{figure=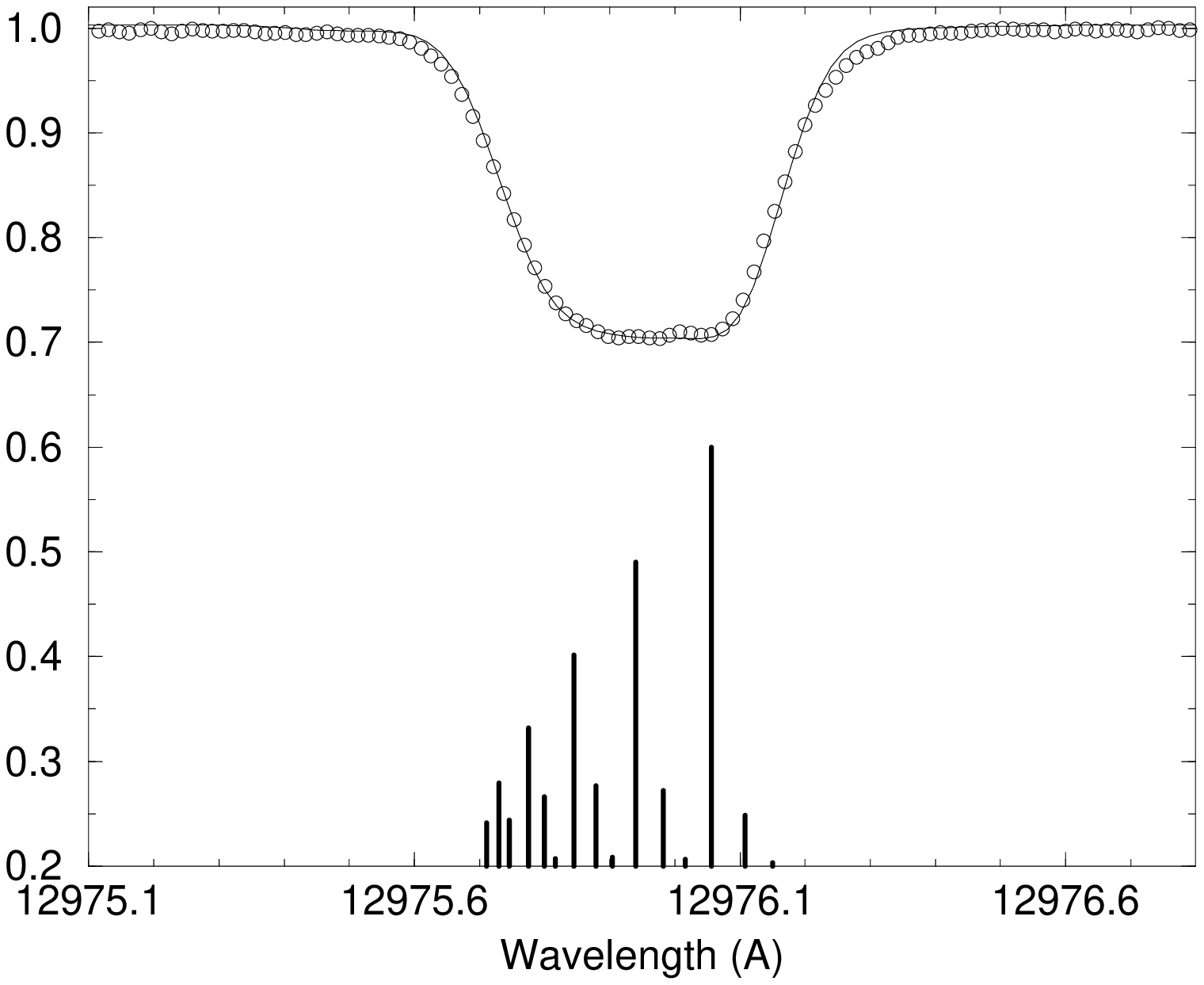,width=7cm}} {\psfig{figure=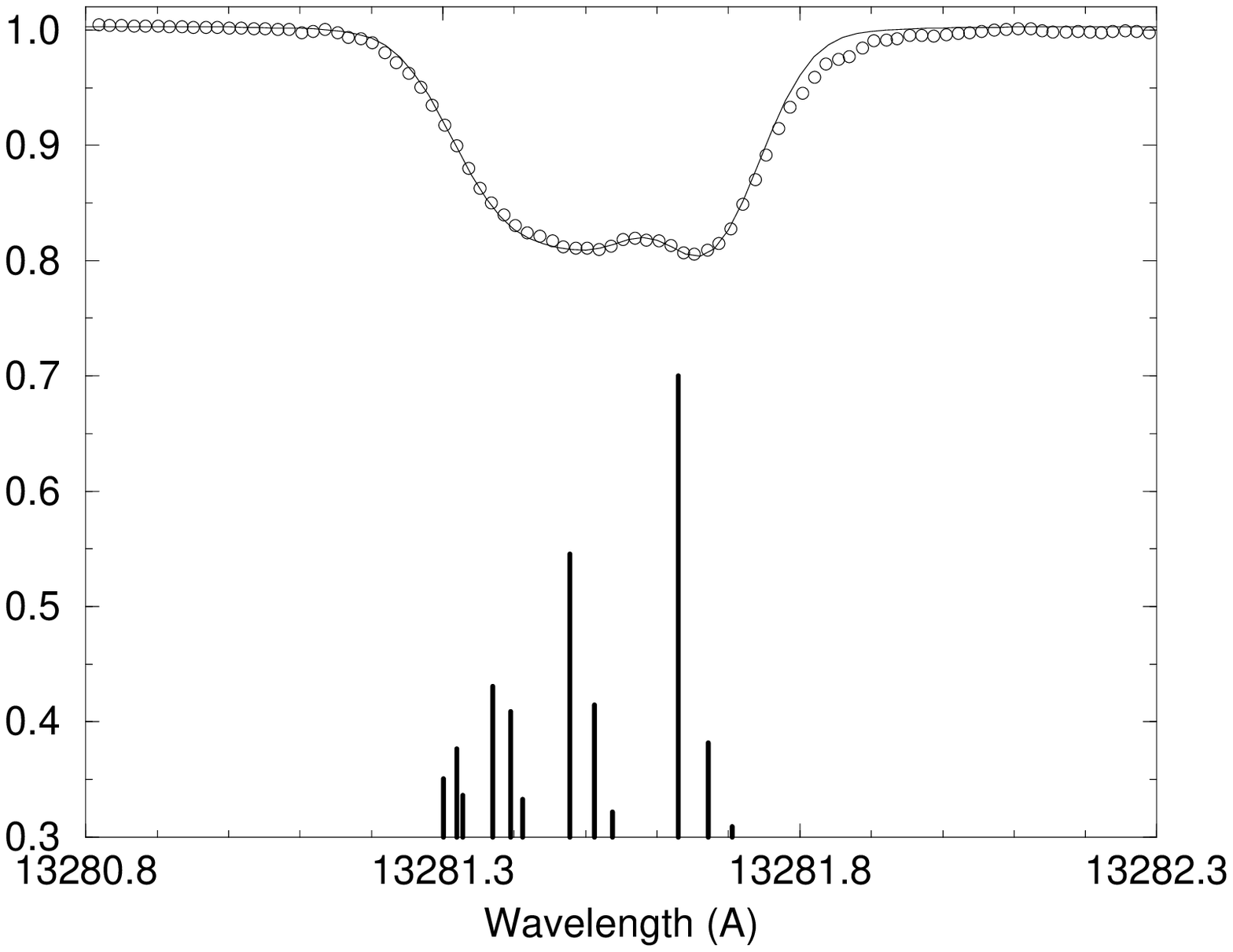,width=7cm}} }
\caption{Fittings for the 12976- and 13281-{\AA} lines. The HFS components were computed 
with corrected experimental interaction constants (Section 3.2). Symbols as in Fig. 1.} 
\end{figure}

\begin{figure}
{ {\psfig{figure=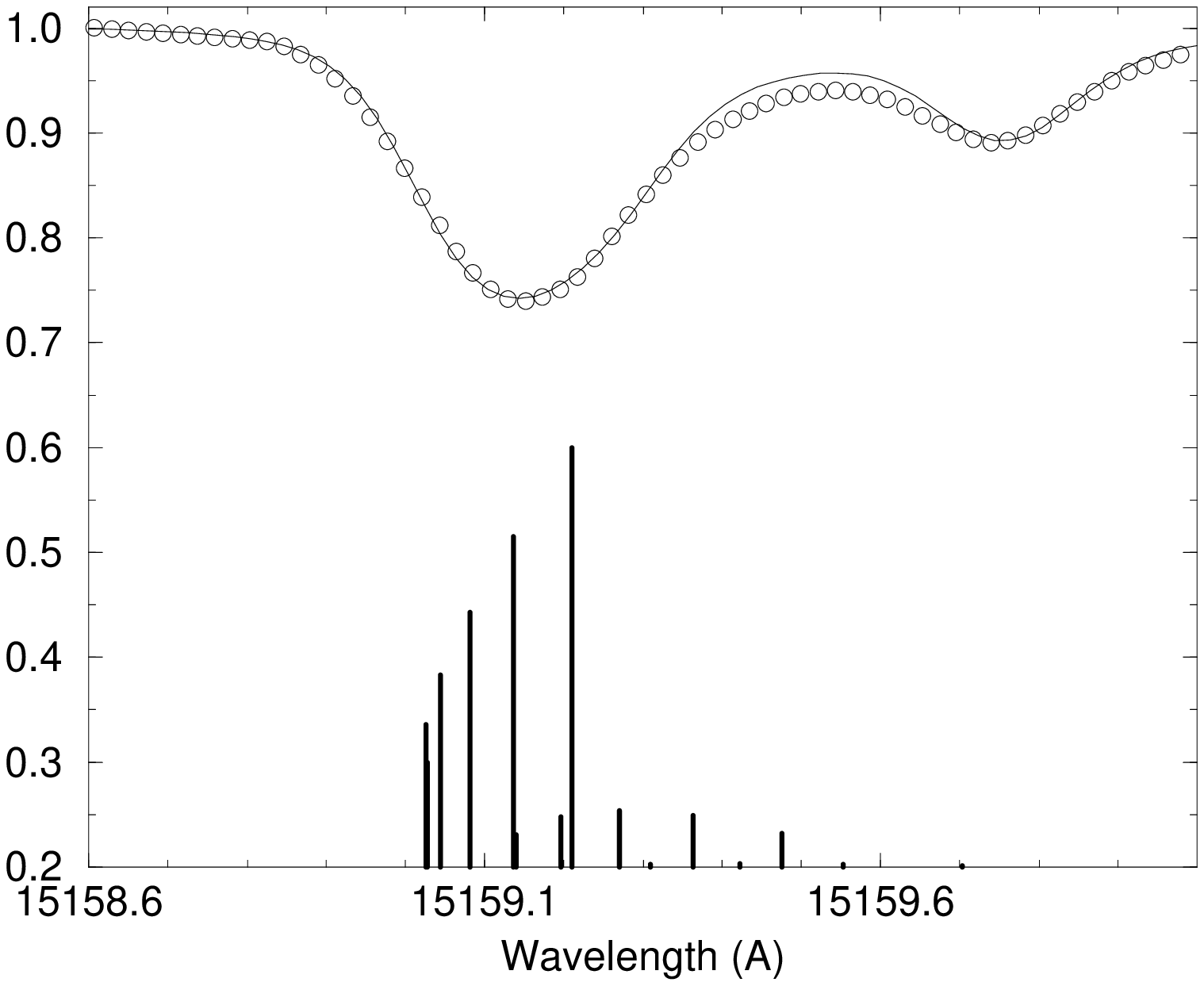,width=7cm}} {\psfig{figure=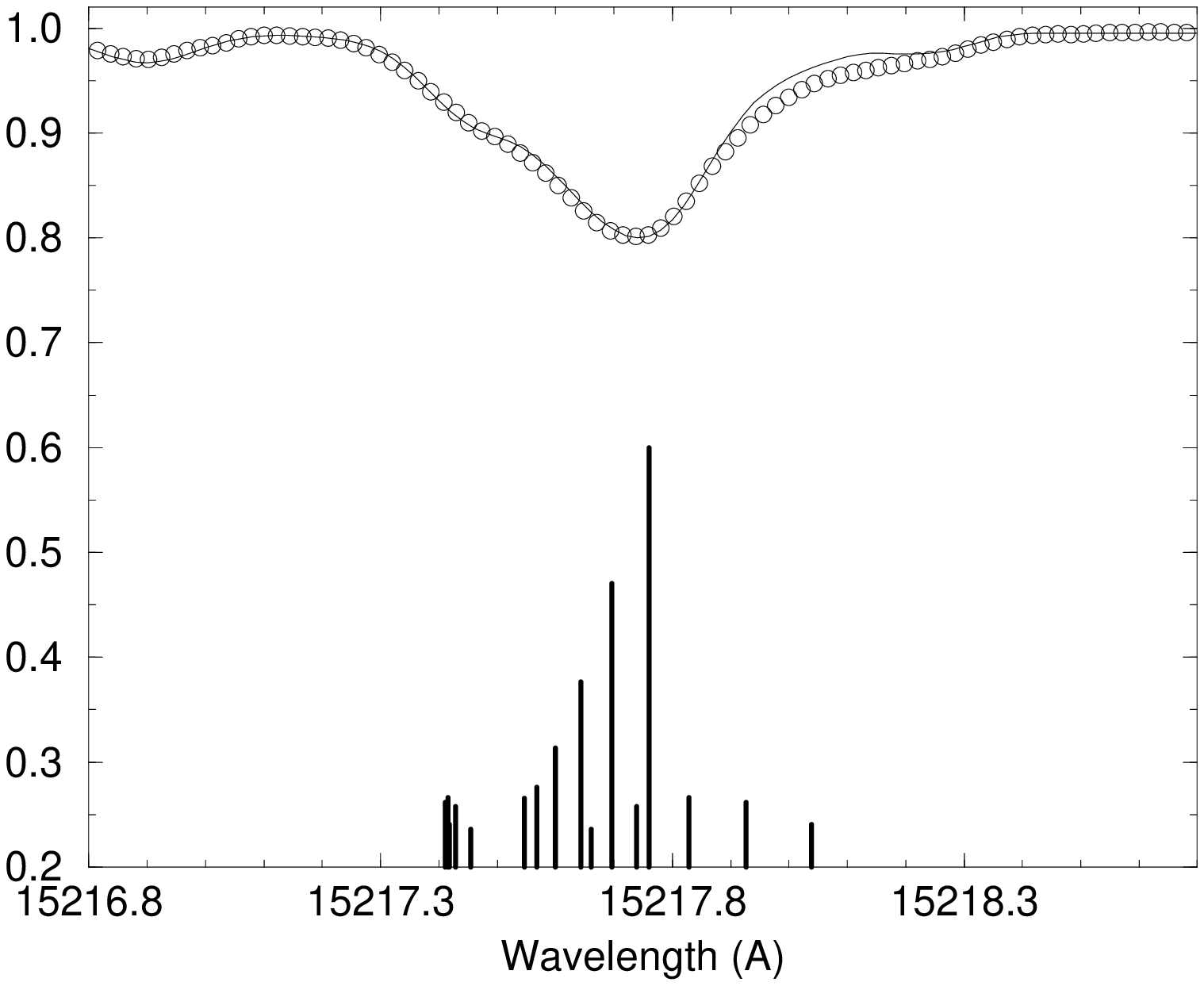,width=7cm}} 
{\psfig{figure=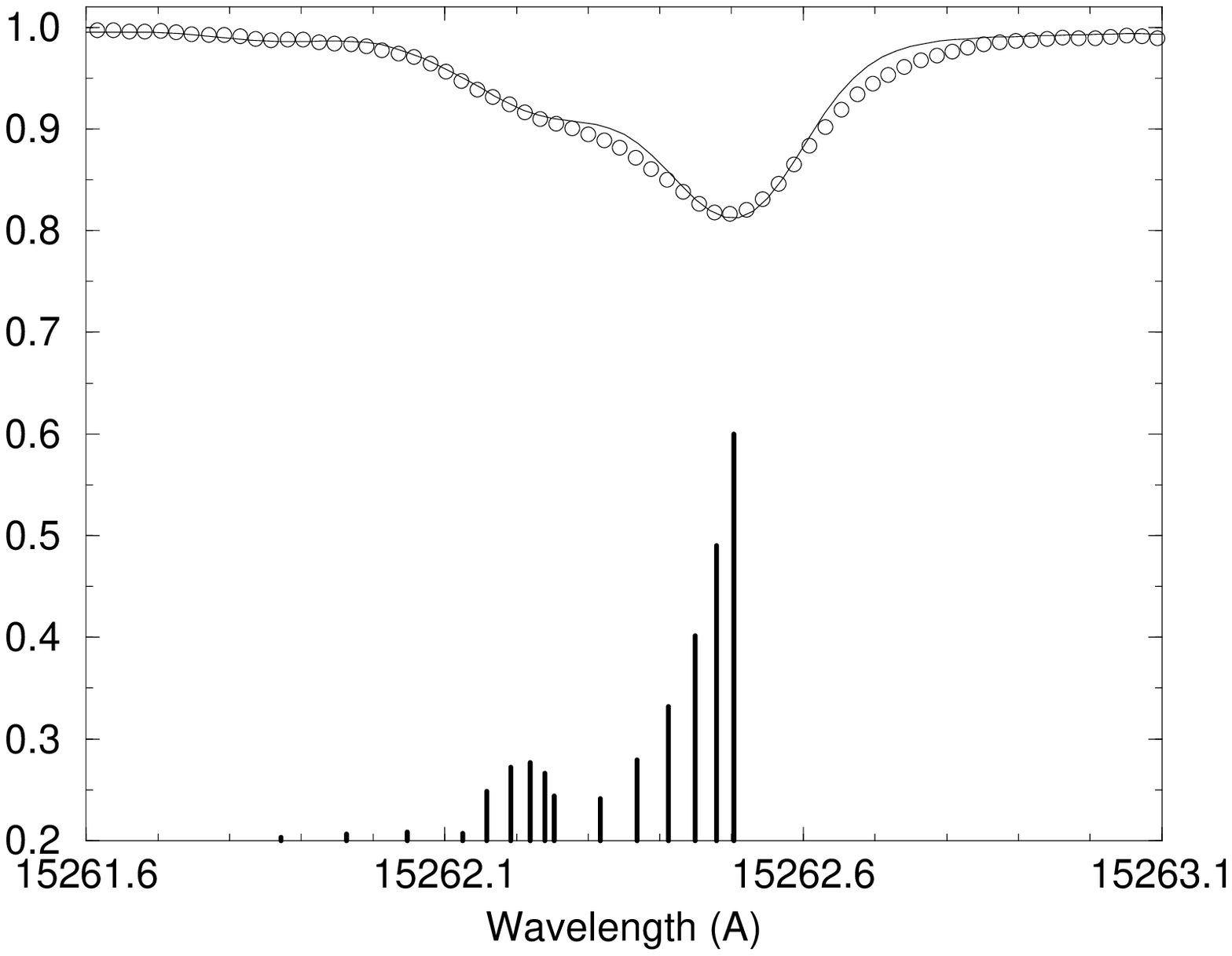,width=7cm}} }
\caption{Fittings for the 15159-, 15217- and 15262-{\AA} lines. The HFS components were computed 
with corrected theoretical interaction constants (Section 3.3). Symbols as in Fig. 1.} 
\end{figure}

\end{document}

\bsp

\label{lastpage}

\end{document}